\documentclass[aps,prx,twocolumn,showpacs,groupedaddress]{revtex4}
\usepackage{graphicx}
\usepackage{color}
\usepackage{amsmath}
 \usepackage{setspace} 
\include{bibliography}

\usepackage{bm}        
\usepackage{amssymb}   
\usepackage{array}

\usepackage[bookmarks=false,linkcolor=blue,urlcolor=blue,colorlinks,citecolor=blue]{hyperref}

\begin{document}

\title{The Mott-Ioffe-Regel limit and resistivity crossover in a tractable electron-phonon model}
\date{\today}
\author{Yochai Werman}
\author{Erez Berg}
\affiliation{Department of Condensed Matter Physics, Weizmann Institute of Science, Rehovot 7610001, Israel}

\begin{abstract}
Many metals display resistivity saturation - a substantial decrease in the slope of the resistivity as a function of temperature, that occurs when the electron scattering rate $\tau^{-1}$ becomes comparable to the Fermi energy $E_F/\hbar$ (the Mott-Ioffe-Regel limit). At such temperatures, the usual description of a metal in terms of ballistically propagating quasiparticles is no longer valid. We present a tractable model of a large $N$ number of electronic bands coupled to $N^2$ optical phonon modes, which displays a crossover behavior in the resistivity at temperatures where $\tau^{-1}\sim E_F/\hbar$. At low temperatures, the resistivity obeys the familiar linear form, while at high temperatures, the resistivity still increases linearly, but with a modified slope (that can be either lower or higher than the low-temperature slope, depending on the band structure). The high temperature non-Boltzmann regime is interpreted by considering the diffusion constant and the compressibility, both of which scale as the inverse square root of the temperature.
\end{abstract}

\maketitle


\emph{Introduction.--} Many materials, especially transition metals and transition metal compounds, display resistivity saturation~\cite{FiskLawson,FiskWebb,Allen1981, savvides1982electrical} - a substantial decrease in the slope of the resistivity as a function of temperature. This reduction in slope occurs in the regime where the experimentally measured lifetime, deduced from the Drude form of the conductivity, approaches the bound $\tau(T)\sim {\hbar}/{E_F}$ (where $E_F$ is the Fermi energy). The 
resistivity often levels off at a value close to the Mott-Ioffe-Regel (MIR) limit~\cite{Ioffe}, $\rho_{\mathrm{MIR}} \sim  h / e^2 k_F$, where $k_F$ is the Fermi momentum, and the crossover occurs at temperatures of a few hundred degrees Kelvin. This phenomenon is believed to be intimately linked to the existence, or lack, of coherent quasiparticles. Despite theoretical progress (for a review, see~\cite{Gunnarsson}), a simple and general explanation is still lacking. 



In systems which exhibit long lived quasiparticles, i.e. the quasiparticle lifetime obeys $\tau\gg \hbar/{E_F}$, the conductivity is well described by the semiclassical Drude theory, resulting in~\cite{Ashcroft}
\begin{eqnarray}
\sigma\propto e^2 E_F\tau k_F^{d-2}
\end{eqnarray}
with $-e$ the electron charge, $k_F$ the Fermi momentum and $d$ the dimensionality, and we set $\hbar=k_B=1$ throughout. However, strong interactions (e.g. electron-phonon scattering) may decrease the lifetime $\tau$ such that the quasiparticles are no longer well defined, and the semiclassical theory is not applicable. The absence of an alternative, well-controlled theoretical description has made further progress difficult. Numerical studies and theoretical arguments~\cite{chakraborty1979boltzmann, Gunnarsson,Calandra,Calandra2,Millis, Allen1981,auerbach1984universal, Allen, Allen2002, Gurvitch, Gunnarsson1,Christoph} indicate that for electron-phonon systems, the resistivity may saturate under certain conditions.

\begin{figure}[h!]
\includegraphics[width=0.5\textwidth]{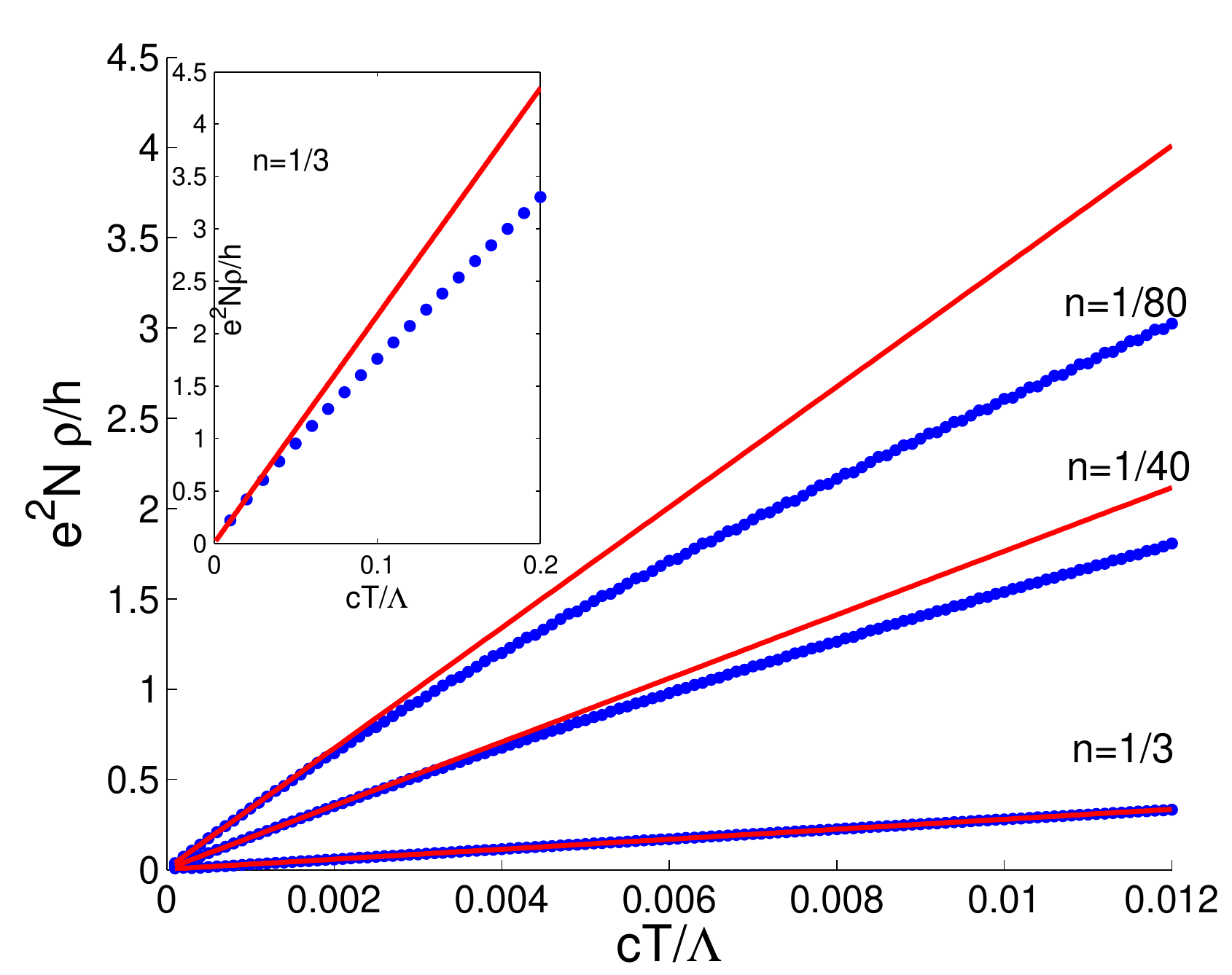}
  \caption{(Color online) Resistivity 
  as a function of $cT/\Lambda$, where $\Lambda$ is the bandwidth, for a two-dimensional tight binding model on a square lattice. The resistivities are calculated 
 for filling $n=1/80, 1/40 ,1/3$ per electronic flavor. The red lines are fits to the low-temperature linear behavior. The $n=1/3$ resistivity continues to rise linearly throughout the shown temperature regime, and crosses over to non-linear behavior only at temperatures comparable to $\Lambda/c$, as shown in the inset.}
\label{fig:ressat}
  \centering
    \end{figure}


In this work, we present a simple, tractable electron-phonon model displaying a crossover behavior in the resistivity at temperature where $E_F\tau\sim1$. 
At the corresponding temperature, the low-$T$ quasiparticle transport crosses over to a distinctly non-Boltzmann behavior.

Some transition metals that display resistivity saturation, such as the A15 compounds~\cite{FiskWebb, Calandra,Gunnarsson1}, are characterized by strong electron-phonon coupling. In these systems, all the 5$d$ orbitals participate in the electronic states near the Fermi energy. 
Motivated by these features, we introduce a model of $N$ identical electronic bands interacting with $N^2$ flavors of optical phonons, with an interaction strength parametrized by a dimensionless coupling constant $c$; this model is analytically solvable in the limit $N\rightarrow\infty$. Our main result appears in Figure~\ref{fig:ressat}, 
showing the resistivity as a function of temperature to lowest order in $\frac{1}{N}$. 

At low temperatures, $T\ll E_F/c$, 
the resistivity obeys the simple Drude formula $\rho=\frac{2\pi}{e^2}\frac{cT}{\nu v_F^2}$, with $\nu$ the density of states at the Fermi energy and $v_F$ the Fermi velocity. { As the temperature becomes of the order of $E_F/c$, the resistivity starts to deviate from this linear law.} Asymptotically, at  temperatures much higher than $\frac{\Lambda}{c}$ (in the vicinity of the MIR limit), with $\Lambda$ the bandwidth, the resistivity is again linear in the temperature, but the slope is modified by a factor of $\frac{v_F^2}{{2}\color{black}\left\langle v^2\right\rangle}$, where $\left\langle v^2\right\rangle$ is the average of the velocity squared over the entire band. Thus, for $v_F^2\ll\left\langle v^2\right\rangle$, 
this model displays resistivity saturation (i.e., there is a substantial decrease in the slope as the resistivity approaches the MIR limit), while for $v_F^2 >2\left\langle v^2\right\rangle$ 
the resistivity slope increases. 

\emph{Model.--} We investigate a system of $N\gg1$ electron flavors interacting with $N^2\gg N$ optical, dispersionless phonon modes. In this type of large-$N$ expansion, inspired by the work of Fitzpatrick et al.~\cite{Raghu}, the phonon modes act as a momentum and energy bath for the electrons. The Lagrangian of the system is given by
\begin{eqnarray}
&&\mathcal{L}=\sum_{a,n}\int \frac{d^dk}{(2\pi)^d} c^\dagger_a(i\nu_n, \mathbf{k})\left(i\nu_n-\xi_\mathbf{k}\right)c^{\vphantom{\dagger}}_a(i\nu_n,\mathbf{k})\nonumber\\
&&+\frac{1}{2}\sum_{n}\sum_{a,b}\int \frac{d^dk}{(2\pi)^d}\left(K+M\omega_n^2\right)|X_{ab}(\omega_n,\mathbf{k})|^2\nonumber\\
\nonumber\\
&&+\frac{\lambda}{\sqrt{N}}\sum_{a,b,n,m}\int \frac{d^dk}{(2\pi)^d}\frac{d^dq}{(2\pi)^d}X_{ab}
(\omega_n,\mathbf{q})\times\nonumber\\
&&\left(c^\dagger_a(i\nu_m,\mathbf{k})c^{\vphantom{\dagger}}_b(i\nu_m+i\omega_n,\mathbf{k}+\mathbf{q})+a\leftrightarrow b\right),
\label{eq:model}
\end{eqnarray}
where $c^\dagger_a(i\nu_n,\mathbf{k})$ creates an electron of flavor $a=1\dots N$ with momentum $\mathbf{k}$ and Matsubara frequency $i\nu_n$, with an energy $\xi_{\mathbf{k}}=\epsilon_\mathbf{k}-\mu$, with $\epsilon_\mathbf{k}\in[0,\Lambda]$ being an even function of $k$, and $\mu$ is the chemical potential. { In our calculations, we have used a two dimensional tight binding model on a square lattice, with dispersion $\epsilon(\mathbf{k})= - 2 t [\cos(k_x)+\cos(k_y) - 2]$ where $t = \Lambda/8$.}
$X^\dagger(\omega_n,\mathbf{k})$ is the Fourier transform of an $N\times N$ symmetric matrix of phonon displacement operators, with spring constant $K=M\omega_0^2$ and mass $M$. {$\lambda/\sqrt{N}$ is the coupling constant between the displacement of a single phonon and the electrons.}

We consider temperatures much larger than the phonon frequency, $T\gg\omega_0$, and our results are accurate to lowest order in $\frac{\omega_0}{T}$. While much larger than the phonon energy scales, the temperature is still much lower than the Fermi energy, $T\ll E_F$, so that the Fermi occupation numbers are well described by the Heavyside function. 

We define the dimensionless electron-phonon coupling as
\begin{eqnarray}
c=\frac{\lambda^2\nu}{M\omega_0^2}
\end{eqnarray}
with $\nu$ the density of states at the Fermi level. $c$
may be large, so that although $T\ll E_F$, $cT$ may be larger than $E_F$. 
We are thus able to access the regime where the quasiparticle scattering rate is larger than its energy, while keeping the electrons degenerate. Note that although $c$ may be large, the coupling to individual phonon modes is scaled by a factor of $1/\sqrt{N}$ (see Eq.~{\ref{eq:model}}). Strong coupling effects, such as lattice instabilities and polaron formation, are thus suppressed in the large $N$ limit (see discussion below). Therefore, the large $N$ limit allows us to access the regime of strong scattering, $E_F \tau \gtrsim 1$, while avoiding lattice instabilities. 

\begin{figure}	
\includegraphics[width=0.5\textwidth]{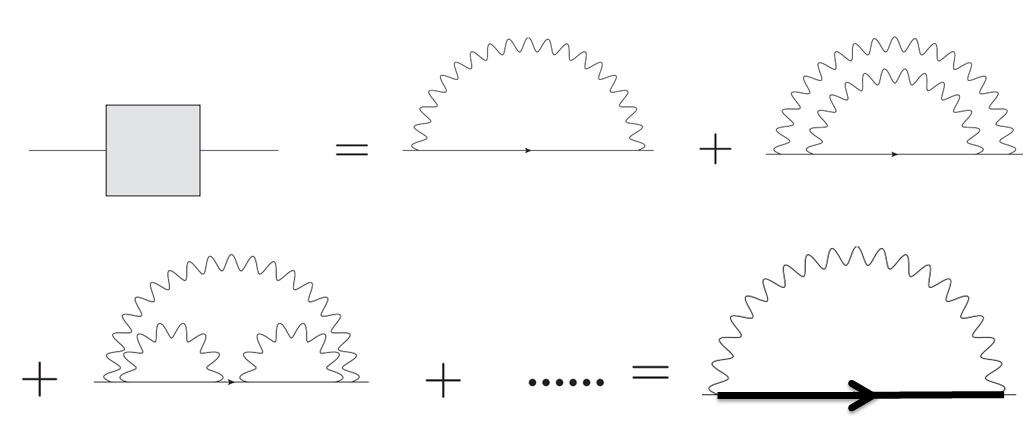}
  \caption{To lowest order in $1/N$, the full set of rainbow diagrams contributes to the electron self energy, denoted by the blue square (color online). The arrows represent bare electron propagators, squiggly lines phonon propagators, and the thick arrow the fully dressed electron propagator.}
\label{fig:rainbow}
  \centering 
\end{figure}

\emph{Results.--} Taking the limit $N\rightarrow \infty$ allows us to solve the model (\ref{eq:model}) order by order in $1/N$. 
Just as in \cite{Raghu}, the full set of rainbow diagrams, depicted in Figure~\ref{fig:rainbow}, contributes to the electron self-energy to lowest order in $1/N$. This results
in a self-consistent Dyson's equation for the fermion self-energy:
\begin{eqnarray}\label{eq:selfenergy}
\Sigma({\omega})=-\frac{cT}{\nu}\int\frac{d^dk}{(2\pi)^d}\frac{1}{{-\omega} + \xi_{\mathbf{k}}+\Sigma(\omega)}.
\label{eq:sigma}
\end{eqnarray}
{ Note that, because of the momentum independent electron-phonon coupling in our model, the self-energy is momentum independent.}

\begin{figure}
 \includegraphics[width=0.5\textwidth]{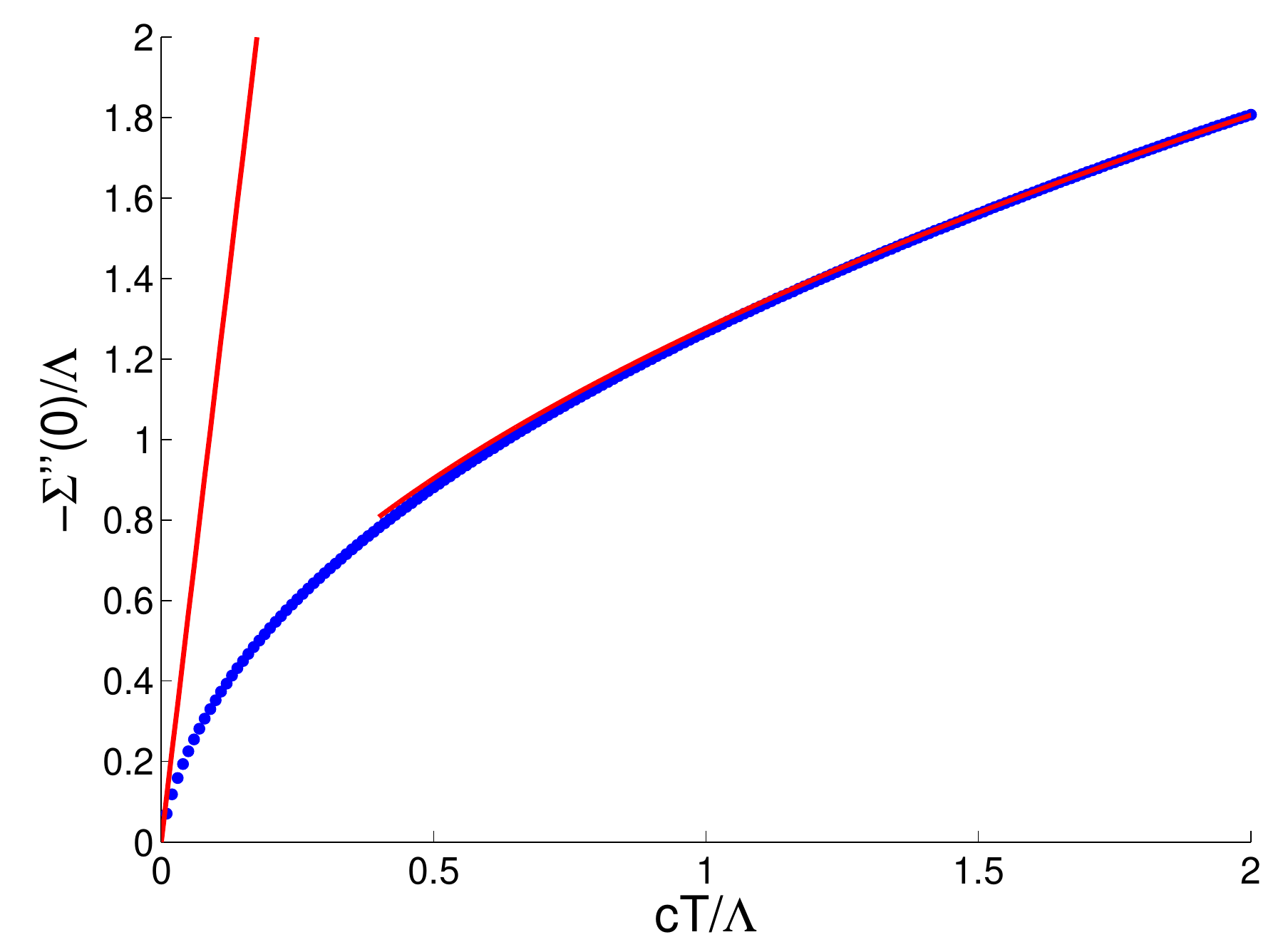}
  \caption{(Color online.) The imaginary part of the self energy at the Fermi level. At low temperatures, $cT\ll E_F$, the self energy increases linearly with temperature, in accordance with the semiclassical result; the linear red line represents a fit to the low-$T$ behavior. At higher temperatures, $\Sigma''(0)$ becomes proportional to $\sqrt{T}$; the second red line is proportional to $\sqrt{T}$, and almost exactly coincides with the numerics. These results are calculated for a two dimensional tight binding system at $n=0.51$.}
\label{fig:selfenergy}
  \centering
\end{figure}

{Eq.~(\ref{eq:sigma}) can be solved explicitly in two asymptotic limits.} In the limit of low temperature, $cT\ll E_F$, the imaginary part of the self energy, $\Sigma''(0)$, coincides with the familiar result, $-\Sigma''(0) \equiv 1/\tau=\pi cT$~\cite{Ziman2, Prange, Mackenzie}. In the other extreme limit, $cT\gg\Lambda$, the scattering rate approaches the asymptotic form $1/\tau=\sqrt{\frac{cT\Omega }{\nu}}$, with $\Omega=\int d^dk/(2\pi)^d$. This result is similar to that found numerically in Ref.~\cite{Millis} for an $N=1$ electron-phonon system, using dynamical mean-field theory (DMFT). The self-energy, obtained from solving Eq.~(\ref{eq:selfenergy}), is shown as a function of temperature in Fig.~\ref{fig:selfenergy}.

\begin{figure}
\includegraphics[width=0.4\textwidth]{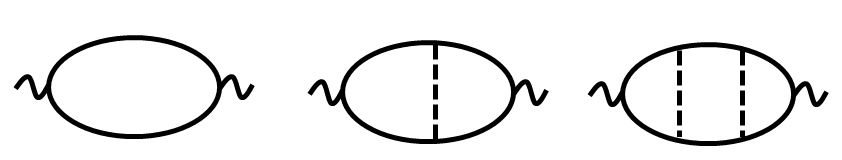}
  \caption{Diagrams contibuting to the conductivity to lowest order in $1/N$. The dashed lines are phonon propagators, while the full lines are fully dressed electron Green's functions. The squiggly lines represent the current operator, adding a factor of $ed\epsilon/dk$.}
\label{fig:ladder}
\end{figure}

Next, we calculate the current-current correlation function, defined as
\begin{eqnarray}
\Pi(i\omega_n,T)&=&\int\frac{d^dk}{(2\pi)^d} \left\langle \vec{J}(\mathbf{k},i\omega_n) \vec{J}(-\mathbf{k},-i\omega_n)\right\rangle, \mbox{with }\nonumber\\
 \vec{J}(\mathbf{k},i\omega_n)&=& \frac{ev_\mathbf{k}}{\beta}\sum_{a,m} c^\dagger_a(i\nu_m,\mathbf{k})c_a(i\nu_m-i\omega_n,\mathbf{k}).\nonumber\\
\end{eqnarray}
Here, $v_\mathbf{k} = \partial\epsilon_\mathbf{k}/{\partial\mathbf{k}}$. To lowest order in $1/N$, the conductivity is composed of ladder diagrams of the form shown in Figure~\ref{fig:ladder}, consisting of non-crossing phonon propagators and fully dressed electron Green's functions. Since the electron dispersion is an even function of momentum and the phonons are dispersionless, all vertex corrections vanish to this order, and we are left with
\begin{eqnarray}\label{eq:conductivity}
\sigma(T)&=&\lim_{\omega\rightarrow 0}\frac{\Im\Pi(\omega_n\rightarrow \omega+i\delta,T)}{\omega}\nonumber\\
&=&\lim_{\omega\rightarrow 0}\frac{e^2N}{\beta\omega}\Im\sum_{\nu_n}\int\frac{d^dk}{(2\pi)^d} v_\mathbf{k}^2 \nonumber\\
&&\times \mathcal{G}(i\nu_n,\mathbf{k})\mathcal{G}(i\nu_n+i\omega_n,\mathbf{k})|_{i\omega_n\rightarrow\omega+i\delta}\nonumber\\
&\approx&e^2N\int\frac{d^dk}{(2\pi)^d}v_\mathbf{k}^2 \left[A(\mathbf{k},\omega=0)\right]^2.
\label{eq:conduct}
\end{eqnarray}
$\mathcal{G}(i\nu_n,k)$ is the fully dressed electron Green's function, and $A(k,\omega)$ is the electron spectral function, $A(k,\omega)=-2\Im\frac{1}{\omega-\xi_{\mathbf{k}}-\Sigma(\omega)}$. In the last line of Eq.~(\ref{eq:conduct}), we have inserted the spectral representation of the Green's function, performed the Matsubara summation over $\nu_n$ (see, e.g.,~\cite{Mahan}), 
and used the fact that the Fermi occupation function $n_F$ obeys $\frac{dn_F(\epsilon)}{d\epsilon} \approx -\delta(\epsilon)$ in the regime $T\ll E_F$, assuming that $A(\mathbf{k},\omega)$ changes slowly on the scale of $T$. This is justified because, at low temperature, $A(\mathbf{k},\omega)$ varies on the scale of $\Sigma''(\omega = 0,T \ll E_F/c) \sim cT$, assumed to be much larger than $T$. [Here we have assumed that the density of states, and hence $\Sigma''(\omega)$, vary slowly around zero energy on the scale of $T$.] At high temperature the spectral function varies on the scale of $\Sigma''(\omega=0,T \gg \Lambda /c) \sim \sqrt{\frac{cT\Omega}{\nu}}\gg T$ [see discussion below Eq.~(\ref{eq:sigma})].

{
To find the conductivity as a function of temperature for a fixed density of electrons (as appropriate for solids), we solve Eq.~(\ref{eq:sigma}) and the equation for the density per flavor
\begin{eqnarray}\label{eq:chempotential}
n &=& \frac{1}{N} \sum_a \int \frac{d^d k}{(2\pi)^2} \langle  c^\dagger_a (\mathbf{k}) c^{\vphantom{\dagger}}_a(\mathbf{k}) \rangle \nonumber \\
&\approx&  \int \frac{d^d k}{(2\pi)^2} \int_{-\infty}^0\frac{d\omega}{2\pi} A (\mathbf{k},\omega)
\label{eq:n}
\end{eqnarray}
simultaneously for $\Sigma(\omega)$ and $\mu$ [where in the last line of Eq.~(\ref{eq:n}) we have used the fact that $T\ll E_F$]. We use Eq.~(\ref{eq:conduct}) to find the conductivity. The resulting resistivity vs. $T$ is plotted in Fig.~\ref{fig:ressat}, for several values of the electronic density $n$. At low temperature, $\rho(T)\propto T$; at higher temperatures, where $T\sim E_F / c$, the resistivity starts curving downward, although it keeps increasing.
}

{ To understand this behavior, it is useful to analyze the resistivity in the asymptotic limits of either low or high temperatures. In the limit $cT\ll E_F$, the Kubo formula gives the familiar result
\begin{eqnarray}
\sigma(T\ll E_F/c)=N\frac{e^2}{2\pi}\frac{\nu v_F^2}{cT}=N\frac{e^2}{2\pi}k_F^{d-2}E_F\tau.
\end{eqnarray}
Note that when $T \sim E_F/c$, $E_F \tau$ becomes of order unity; this is the MIR limit.

In the high temperature regime, $cT\gg\Lambda$, the scattering rate grows as $|\Sigma''(\omega=0)|\approx \sqrt{\frac{cT \Omega }{\nu}}\gg\Lambda$. In this regime, one can solve Eqs.~(\ref{eq:selfenergy}, \ref{eq:chempotential}), as described in the Appendix. This gives
\begin{eqnarray}
\mu(T)&=&f_0\sqrt{\frac{\Omega cT}{\nu}}, \label{eq:asumptotic} \\
\Sigma(\omega) &=& \frac{1}{2}\left(\omega+\mu(T)-i\sqrt{\frac{4cT\Omega}{\nu}-(\omega+\mu(T))^2}\right) \nonumber
\end{eqnarray}
with $f_0\in[-2,2]$ a dimensionless parameter obtained by solving the equation $n=\Omega\int_{-2}^{f_0}\frac{dy}{2\pi}\sqrt{4-y^2}.$} 
Inserting this form of $\mu(T)$ and $\Sigma(\omega)$ into Eq.~(\ref{eq:conduct}), we obtain
\begin{eqnarray}
\sigma(T\gg\Lambda/c)&\approx&N\frac{e^2\nu \left(1-\frac{f_0^2}{4} \right)}{\pi\color{black}cT \Omega}\int\frac{d^dk}{(2\pi)^d} \, v_\mathbf{k}^2\nonumber\\
&=&N\frac{e^2}{2\pi} \left(1-\frac{f_0^2}{4} \right)  \frac{2\color{black}\nu\left\langle v_{\mathbf{k}}^2\right\rangle_{BZ}}{cT},
\end{eqnarray}
where $\langle f_\mathbf{k} \rangle_{BZ} \equiv \frac{1}{\Omega} \int \frac{d^dk}{(2 \pi)^d} f_\mathbf{k} $. 

Thus, at low temperatures, the resistivity increases linearly with temperature according to the familiar semiclassical result $\rho=\frac{2\pi}{Ne^2}\frac{cT}{\nu v_F^2}$; in the high-$T$ regime, the resistivity is given by
\begin{eqnarray}
\rho=\rho_0+\frac{2\pi}{Ne^2}\frac{cT}{2\nu \left\langle v_\mathbf{k}^2\right\rangle_{BZ}}\frac{1}{1-f_0^2/4},
\end{eqnarray}
with $\rho_0$ calculated in the next order in $\Lambda/cT$ to be 
\begin{eqnarray}
\rho_0=\frac{\pi}{Ne^2}\frac{\left\langle v_\mathbf{k}^2 \epsilon_\mathbf{k}^2 \right\rangle_{BZ}}{\Omega\left\langle v_\mathbf{k} ^2\right\rangle_{BZ}^2}\frac{f_0^2}{1-f_0^2/4}.\nonumber\\
\end{eqnarray}


For a typical electronic dispersion, we get $\rho_0 \propto \frac{2\pi}{Ne^2}k_F^{2-d}$, with a proportionality coefficient which is a number of order unity (that depends on the particular band structure). 


Therefore, within our model, the resistivity increases linearly with temperature in both the low and high temperature regimes, but a crossover to a different slope occurs when the resistivity reaches a value of the order of the Mott-Ioffe-Regel limit. Systems with $v_F^2\ll\left\langle v^2\right\rangle_{BZ}$ will exhibit a marked decrease in slope, reminiscent of resistivity saturation. The high-$T$ system displays distinctly non-Boltzmann transport, without well-defined quasiparticles. 

The fact that well-defined quasiparticles no longer exist in the high-$T$ regime motivates a description of the transport in terms of a diffusion constant, utilizing the Einstein relation~\cite{Hartnoll}, which relates the conductivity to the compressibility and to the diffusion constant. In order to find the diffusion constant, we calculate the compressibility
\begin{eqnarray}
&&\chi= N \frac{\partial n}{\partial \mu}=N\times\\
&&\left[ \int \frac{d^dk}{(2\pi)^d} A(\mathbf{k},\omega=0) + \int d\epsilon n_F(\epsilon) \int \frac{d^dk}{(2\pi)^d} \frac{d A(\mathbf{k},\epsilon)}{d\mu}\right] \notag.
\end{eqnarray}
In the low-$T$ limit, the compressibility is given by 
$\chi\propto\nu$, independent of temperature. In the high-$T$ limit, using Eq.~(\ref{eq:asumptotic}), $A(\mathbf{k},\omega)= \frac{4\nu}{cT\Omega}\sqrt{\frac{cT\Omega}{\nu}-(\omega+\mu)^2}$, and we find that $\chi\sim N\sqrt{\frac{\Omega}{\frac{c}{\nu}T}}$.  Using the Einstein relations, we deduce
\begin{eqnarray}
D=\frac{\sigma}{\chi}\sim\left\{
\begin{array}{rl}
\frac{v_F^2}{cT}& \text{if } cT\ll E_F,\\
\frac{\left\langle v^2\right\rangle}{\sqrt{\frac{c}{\nu}\Omega T}} & \text{if } cT\gg\Lambda.
\label{eq:diffusion}
\end{array} \right.
\end{eqnarray}

We assume that the crossover between these results occurs at $cT\sim E_F$, as it does for the conductivity.	

\emph{Discussion.--} 
Our model exhibits a crossover from a 
low-temperature regime where the transport can be thought of in terms of long-lived ballistic quasi-particles, to a 
high-temperature regime where the inverse lifetime of a momentum state is comparable to its energy. In the high temperature 
regime, since the momentum of the electron system is not even approximately conserved, the transport is most naturally thought of in terms of charge diffusion~\cite{Hartnoll}. 
The high temperature behavior of the diffusion constant, Eq.~(\ref{eq:diffusion}), can be understood simply as follows. 
The physics of each site is governed by an $N\times N$ Hamiltonian of phonon deformation potentials operating in the Hilbert space of the $N$ electron flavors. The matrix element of the site Hamiltonian are random, taken from the thermal Gaussian distribution of width $\Delta E=\lambda\sqrt{\frac{T}{K}}=\sqrt{\frac{c}{\nu}T}$. According to random matrix theory, the distribution of eigenenergies of the single site Hamiltonian is given by the Wigner semicircle law, such that the density of states is $\rho(\epsilon)=\frac{N}{2\pi\Delta E^2}\sqrt{4(\Delta E)^2-\epsilon^2}$
~\cite{Forrester}. The maximum momentum relaxation rate at high temperatures is of the order of the inverse bandwidth, hence  $\tau^{-1}\propto\sqrt{T}$. 

The diffusion constant is 
composed of a typical velocity squared multiplied by a time scale; at high temperatures it is then simply given by
\begin{eqnarray}
D\sim{\left\langle v^2\right\rangle}_{BZ} \tau=\frac{\left\langle v^2\right\rangle_{BZ}}{\Delta E},
\end{eqnarray}
which is the behavior that appears in Eq.~(\ref{eq:diffusion}). The averaged band velocity is used, rather than the Fermi velocity; at high temperature, the entire band participates in the transport. In the high temperature regime, the compressibility scales as $N/\Delta E$, and therefore $\chi(cT\gg\Lambda)\propto\frac{1}{\sqrt{cT/\nu}}$. Using these simplified relations for the diffusion constant and the compressibility, the conductivity is given by $\sigma=D\chi$, resulting in the two asymptotic behavior obtained above.



Hartnoll~\cite{Hartnoll} postulated a bound on the diffusion constant,
\begin{eqnarray}
D\ge\frac{v_F^2}{T}.
\label{eq:bound}
\end{eqnarray}
In our system, this bound is violated at high temperature, as $\Delta E\gg T$ in the regime we consider. This is because the bound of Eq.~(\ref{eq:bound}) is based on an energy-time uncertainty relation for the velocity degredation time, assuming that the typical available energy is given by the temperature $T$. In our system, the phonons can be regarded as a bath that allows dissipation of energy and momentum of the electronic system; there is an additional energy scale $\Delta E$, that corresponds to the rate of dissipation due to coupling to the bath. 


Conspicuously absent from the physics described above are strong coupling effects, namely Anderson localization and polaron physics. In the limit $\omega_0\rightarrow 0$ the phonons 
act as static disorder; 
one may expect that at sufficiently high temperatures, all the electronic states for a given phonon configuration may become localized, which would cause a rapid increase in the resistivity~\cite{Gunnarsson}. However, in the $N\rightarrow\infty$ limit, the intra-site level spacing scales as $\Delta E/N$, and is hence always much smaller than the hopping matrix element between two adjacent sites. Therefore, the physics of Anderson localization does not appear to leading order in $1/N$. 

Similarly, polaronic physics will not appear in the limit $N\rightarrow\infty$. Upon fixing $N$ and upon increasing the electron-phonon coupling constant, the electronic effective mass is enhanced by a Franck-Condon factor that scales as $e^{-\frac{\lambda^2}{2NK\omega_0}}$. Clearly, polaron formation is absent to leading order in $1/N$.

In this sense, the large-$N$ limit is a weak-coupling theory, with the normalized coupling constant
\begin{eqnarray}
\tilde{\lambda}=\frac{\lambda}{\sqrt{N}}.
\end{eqnarray}
Effects which depend on the coupling of an electron to a single phonon, such as localization due to polaron formation, are absent in the large-$N$ limit; on the other hand, non-coherent effects such as momentum degradation, which result from the coupling of an electron to $N^2$ phonons, are not suppressed.

\emph{Conclusions.--} 
We have presented a tractable model of a large number $N$ of electronic flavors coupled to $N^2$ optical phonon modes. The resistivity of this system as a function of temperature exhibits a crossover when the quasi-particle scattering rate becomes of the order of the Fermi energy, from a linear Boltzmann form to a non-Boltzmann regime, where the resistivity is again linear but with an altered slope; depending on the bandstructure and chemical potential, this system may exhibit a reduction of slope (similar to ``resistivity saturation''), or, conversely, an increase in slope at high temperatures. 

While it is satisfying that the crossover behavior at the Mott-Ioffe-Regel limit, $E_F \tau \sim 1$, arises naturally from our model, it is not entirely clear why above the MIR limit the resistivity does not universally saturate; the high--temperature behavior is rather subtle, depending on the shape of the band and the coupling constant. This may, of course, be an artefact of the large--$N$ limit; it could, alternatively, point to the possible importance of the physics missing in our model, such as phonon non-linearity or Coulomb interactions.


\emph{Acknowledgements.--} We thank E. Altman, S. Hartnoll, S. Kivelson, and S. Raghu for useful discussions. This work was supported by the Israel Science Foundation (ISF) under grant 1291/12, by the US-Israel Binational Science Foundation (BSF) under grant 2012079, and by a Marie Curie CIG grant.

\bibliography{paper}

\appendix*

%

\section{Chemical potential at constant density}
\label{sec:chempotential}
In order to calculate the temperature dependence of the chemical potential at high-$T$, we use the ansatz
\begin{eqnarray}
\mu(T\gg\Lambda/c)=f_0\sqrt{\frac{\Omega cT}{\nu}}
\end{eqnarray}
We then calculate the self-energy
\begin{eqnarray}
\Sigma(\omega)=\frac{cT}{\nu}\int \frac{d^dk}{(2\pi)^d}\frac{1}{\omega-\epsilon_k+\mu(T)-\Sigma(\omega)}\approx\nonumber\\
\frac{cT\Omega}{\nu}\frac{1}{\omega+\mu(T)-\Sigma(\omega)}
\end{eqnarray}
which produces
\begin{eqnarray}
\Sigma(\omega)=\frac{1}{2}\sqrt{\frac{cT\Omega}{\nu}}\left(\tilde{\omega}+f_0-i\sqrt{4-(\tilde{\omega}+f_0)^2}\right)\nonumber\\
A(k,\omega)\approx \frac{\sqrt{4-(\tilde{\omega}+f_0)^2}}{\sqrt{\frac{cT\Omega}{\nu}}}\Theta\left(4-(\tilde{\omega}+f_0)^2\right),
\end{eqnarray}
with $\tilde{\omega}=\omega/\sqrt{\frac{cT\Omega}{\nu}}$.
We then self-consistently calculate the electron density, which is given by
\begin{eqnarray}
n=\Omega\int_{-\infty}^0\frac{d\omega}{2\pi}A(\omega)=\Omega\int_{-2}^{f_0}\frac{d\tilde{\omega}}{2\pi}\sqrt{4-\tilde{\omega}^2},
\end{eqnarray}
independent of temperature.

The conductivity is then calculated according to formula \ref{eq:conductivity}, which results in the modification of the high-$T$ resistivity slope by a factor of $1/\left(1-\left(\frac{f_0}{2}\right)^2\right)$.
\end{document}